\documentclass[pre,longbibliography,footinbib,twocolumn,preprintnumbers,amsmath,amssymb]{revtex4-1}

\usepackage{amssymb}
\usepackage{amsmath}

\usepackage{color}
\usepackage[pdftex]{graphicx}

\usepackage{empheq}

\usepackage{mathtools}

\usepackage{makecell}

\usepackage{enumerate}
\usepackage{mathtools}
\usepackage{stmaryrd}

\usepackage{enumitem}

\usepackage{textcomp}
\usepackage{gensymb}

\newcommand{\C}[0]{C}

\newcommand{\dt}[0]{\Delta t}

\newcommand{\kB}[0]{k_{\mathrm{B}}}
\newcommand{\lbox}[0]{\ell_{\mathrm{box}}}

\newcommand{\moy}[1]{\left\langle #1 \right\rangle}
\newcommand{\Nc}[0]{N_C}
\newcommand{\nn}[0]{{\hat{n}}}

\newcommand{\pp}[0]{\vec{p}}
\newcommand{\rcut}[0]{r_{\mathrm{cut}}}
\newcommand{\rr}[0]{\vec{r}}

\DeclareMathOperator{\e}{e}

\definecolor{darkblue}{rgb}{0,0,0.6}
\definecolor{darkred}{rgb}{0.6,0,0}
\usepackage[colorlinks=true,urlcolor=darkblue,citecolor=darkblue,linkcolor=darkred]{hyperref}

\usepackage{soul}

\begin{document}

\title{Isotropic active colloids: explicit vs. implicit descriptions of propulsion mechanisms}

 \author{Jeanne Decayeux}
 \affiliation{Sorbonne Universit\'e, CNRS, Physico-Chimie des \'Electrolytes et Nanosyst\`emes Interfaciaux (PHENIX), 4 Place Jussieu, 75005 Paris, France}

 \author{Jacques Fries}
 \affiliation{Sorbonne Universit\'e, CNRS, Physico-Chimie des \'Electrolytes et Nanosyst\`emes Interfaciaux (PHENIX), 4 Place Jussieu, 75005 Paris, France}
 
  \author{Vincent Dahirel}
 \affiliation{Sorbonne Universit\'e, CNRS, Physico-Chimie des \'Electrolytes et Nanosyst\`emes Interfaciaux (PHENIX), 4 Place Jussieu, 75005 Paris, France}
 
  \author{Marie Jardat}
 \affiliation{Sorbonne Universit\'e, CNRS, Physico-Chimie des \'Electrolytes et Nanosyst\`emes Interfaciaux (PHENIX), 4 Place Jussieu, 75005 Paris, France}
 
  \author{Pierre Illien}
 \affiliation{Sorbonne Universit\'e, CNRS, Physico-Chimie des \'Electrolytes et Nanosyst\`emes Interfaciaux (PHENIX), 4 Place Jussieu, 75005 Paris, France}

\begin{abstract}
Modeling the couplings between active particles often neglects the possible many-body effects that control the propulsion mechanism. Accounting for such effects requires the explicit modeling of the molecular details at the origin of activity.  Here, we take advantage of a recent two-dimensional model of isotropic active particles whose propulsion originates from the interactions between solute particles in the bath.  The colloid catalyzes a chemical reaction in its vicinity, which results in a local phase separation of solute particles, and the density fluctuations of solute particles cause the enhanced diffusion of the colloid. In this paper, we investigate an assembly of such active particles, using (i)~an explicit model, where the microscopic dynamics of the solute particles is accounted for; and (ii)~an implicit model, whose parameters are inferred from the explicit model at infinite dilution.  In the explicit solute model, the effective diffusion coefficient of the active colloids strongly decreases with  density, an effect which is not captured by the derived implicit model. 
 This suggests that classical models, which usually decouple pair interactions from activity, fail to describe collective dynamics in active colloidal systems driven by solute-solute interactions. 
 \end{abstract}

\date{\today}

\maketitle

\section{Introduction}

During the past decades, self-propelled colloids have become a paradigmatic model to study non-equilibrium transport in complex environments~\cite{Lauga2009,Bechinger2016, Zottl2016a, Illien2017}.
In these systems, a modification of the chemical composition around the surface of the colloidal particle typically generates concentration or electric gradients, which result in an effective propulsion of the particle.
Such mechanisms have been particularly studied (both experimentally and theoretically) in the case of Janus particles, where only a part of the colloidal surface catalyzes the chemical reaction~\cite{Ruckner2007a,Golestanian2007,Ebbens2010,Samin2015,Wurger2015,Volpe2011,Buttinoni2012,Buttinoni2013,Oshanin2017,Jiang2010,Safaei2019,Popescu}. 
Isotropic particles can also self-propel in specific cases, as a consequence of a persistent instability within the  surrounding of the particle~\cite{Golestanian2009, Valeriani2013,Golestanian2019,DeBuyl2013a,Michelin2013,Michelin2014,Izri2014,Herminghaus2014,Hu2019,DeCorato2020,Illien2020,Izzet2020,Farutin2021}. 
A recent original mechanism~\cite{Decayeux2021a,Decayeux2022} explored in our group sheds light on the possible interplay between non-equilibrium condensate formation (or mesoscale phase separation) and enhanced diffusion. More precisely, when metastable droplets of solute particles form in the vicinity of the surface of an isotropic colloid, they generate forces that push the colloid.

The physics of active colloids becomes particularly rich when it comes to their collective properties. The interplay between the structure of the suspension and the activity of the colloidal particles has been addressed in such systems, revealing activity-dependent clustering and phase separation of the active particles themselves (usually called motility-induced phase separation 
\cite{Cates2015}). In the situation where the propulsion of the active particles relies on mescoscale phase separations, it is expected that the latter will be related to the collective properties of the colloids at larger length scales: both effects are interlinked, and connected to dynamical properties.

However, in order to describe the dynamics of a suspension of active colloidal particles, most models rely on an \emph{implicit} description of the molecules surrounding the active colloidal particles, and in particular of the solvent and solute molecules whose interaction with the colloid  generates its propulsion. A classical model is that of Active Brownian Particles (ABPs), in which activity appears through an effective propulsion velocity of fixed amplitude and fluctuating orientation, with a fixed persistence time.
There is a large body of literature on ABPs, and this model has been particularly fruitful to analyze the role of activity in the phase transition of assemblies of active colloids~\cite{Fily2012,Redner2013}.

\begin{figure*}
    \centering
    \includegraphics[width=2\columnwidth]{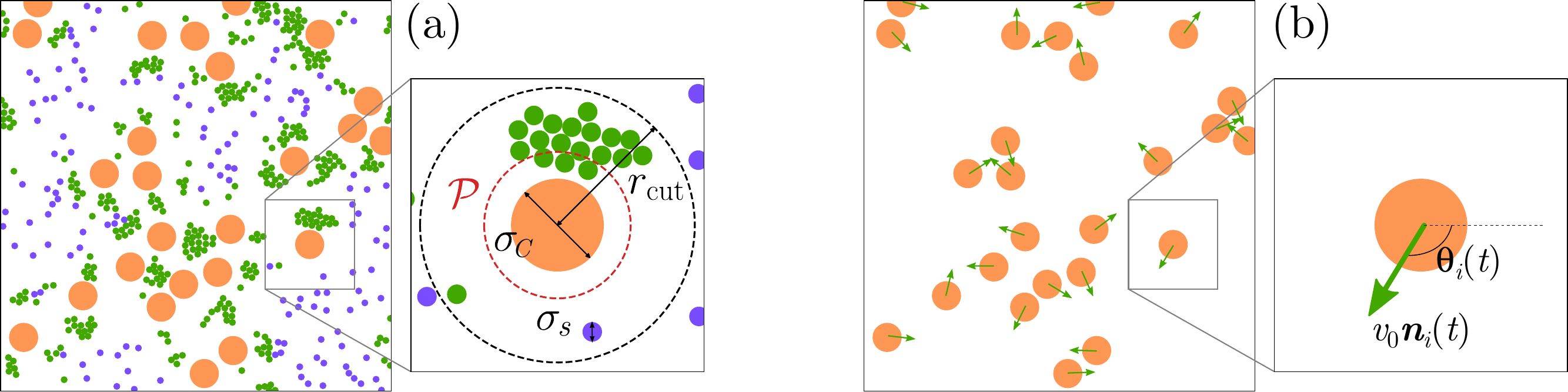}
    \caption{Snapshot of the systems. (a) Simulations with an explicit propulsion mechanism: $\Nc$ colloidal particles, surrounded by $N$ solute particles (see Section \ref{sec:model_explicit}). Particles of type $A$ (purple) interact via a purely repulsive WCA potential, whereas particles of type $B$ (green) interact via an attractive LJ potential. Each colloid triggers the reaction $A  \to B$.  The density fluctuations resulting from the mesophase separation in the vicinity of each colloid lead to their propulsion. (b) Implicit simulations, relying on the Active Brownian Particles (ABP) model. Each particle has a propulsion velocity $v_0 \hat{n}_i(t)$, whose orientation fluctuates (see Section \ref{sec:model_ABP}).}
    \label{fig:snapshot}
\end{figure*}

Modeling suspensions of colloids with implicit descriptions (i.e. where the microscopic details of the solvent and solute are not described by explicit degrees of freedom) has been a central question for \emph{passive} colloids, and relies on hypotheses that have been widely studied~\cite{Hansen_Lowen,Lowen_1993,Likos_2001}. In particular, a seminal question concerns the evolution of effective interactions when the colloidal density increases~\cite{Bolhuis2001,Dobnikar_2002,Dahirel_Hansen}. 

The effective potential may be unambiguously defined as the potential of mean force (PMF) between two infinitely diluted colloidal particles. From a simulation perspective, the mean force may be computed by fixing two colloidal particles in a large enough simulation box, and averaging the force over all solute configurations~\cite{Wu98}.  
One may also define a pair effective potential at finite concentration, by directly inverting the radial distribution function (for instance using HNC inversion methods~\cite{Bolhuis2002} or through a numerical fitting procedure, based on force fitting~\cite{Ercolessi_1994} or on machine learning techniques~\cite{Mishin21}). Nevertheless, if the resulting potential depends on colloidal density, it clearly means that many body effects are important. Such knowledge helps understanding the physics of the system, but the strength of multibody terms questions the validity of the pair additivity assumption. In such cases, the use of effective interactions may no longer be the best solution to define an effective model. One may rather resort to a n-body treatment such as a DFT scheme~\cite{Lowen93}, where the effective free energy is computed as a function of the coordinates of all colloidal particles. 

For the case of active colloids, these questions are known to be particularly challenging~\cite{Liebchen2019,Liebchen2022}, and have been addressed only elusively.  
 The most common simulation techniques for active colloids contain approximations that are not present for passive ones: the forces induced by the solvent are typically separated into passive and active terms which are related to the separation of solvent-averaged induced forces into passive and active terms. 
For example, in the ABP model, the motion of the colloids is governed by a passive component that accounts for the effect of intercolloid conservative forces,  an active term that accounts for the colloid propulsion, and thermal noise. 
A central hypothesis is that the active term is independent from the passive ones. In other words, in the simplest models, the activity of an individual colloid does not depend on how it interacts with its neighbours. As a consequence, the same effective forces as those of passive models of similar colloids may be used, and the active term is that of an infinitely dilute active colloid. 

The use of effective parameters that (i) have been determined at infinite dilution and (ii) are assumed to be independent from one another, may be challenged by a systematic comparison of implicit solute and explicit solute simulations.  
Such comparison has been made several times for passive colloids~\cite{Dahirel_Hansen}, but it has never been proposed for active colloids. 
In the present work, we rely on a model of active isotropic colloid that we recently developed~\cite{Decayeux2021a, Decayeux2022}. The system is driven away from equilibrium under the effect of a chemical reaction, which converts repulsive solute particles into attracting ones. Here, the reacting solutes are described \emph{explicitly} in the model. Therefore, the  activity is not part of an input parameter, but it emerges in the system as a result of the behavior of the interacting solutes. 
The chemical reactions at play are simple unimolecular conversions, whose rate deliberately break detailed balance. In this perspective, we can draw a parallel between the dynamics of our model, which displays mesoscale phase separation, and that of active microemulsions, which recently gained a lot of interest for the modeling of biological condensates~\cite{zwicker2022,Weber2019}. 
In the present article, we run Brownian dynamics simulations of this \emph{explicit solute} model, with a unique colloidal particle in the simulation box, in order to infer the dynamical parameters of an implicit model. We then make a systematic comparison of the properties of (i) our original explicit solute model (Fig. \ref{fig:snapshot}(a)) and (ii) the implicit model (Fig. \ref{fig:snapshot}(b)), as a function of the colloid density. We work in a domain of parameters where activity is not expected to induce phase separation of the colloids.

In the attempt to bridge scales in simple active systems, our results show the failure of a typical coarse-graining strategy, which has been designed for passive colloids. We show that, even if the interaction potential used in the implicit model was inferred from explicit solute simulations, the structure of the suspension differs from that of the explicit model. Moreover, the effective diffusion of colloids strongly depends on the colloid density in the explicit solute model, whereas this dependence is much weaker with the ABP simulations. These limitations in the use of pair effective potentials are clear signatures of many-body effects, which appear to be much stronger for active colloids than in similar passive systems~\cite{Dahirel_Hansen}.

\section{Numerical models}
\label{section-model}

\subsection{Explicit propulsion mechanism}
\label{sec:model_explicit}

We first describe the simulations where the dynamics of the solute particles is  described explicitly. {In earlier related models of active colloids with explicit solute particles~\cite{Yang2014, Reigh2018, Niu2018, Robertson2020}, although interactions within the bath were included, the presence of a structural anisotropy of the colloidal particle reduces the role of solute-solute interactions in the propulsion mechanism. In our model, the solute-solute interactions are accounted for, and their central role on the propulsion mechanism is explored}. This model will be referred to as the \emph{explicit } model. More precisely, we study a two-dimensional system where $\Nc$ colloidal particles of diameter $\sigma_{C}$ and $N$ solute particles of diameter $\sigma_{s}$ are placed in a square box of length $\lbox$ (Fig.~\ref{fig:snapshot}(a)) { in an implicit solvent. We thus consider a system composed of a solvent and of two types of immersed particles, the latter being referred to as colloid and solutes as they differ in size. As detailed hereafter, small immersed particles are actually of two types $A$ and $B$, so that we actually consider a quaternary system.}
For both explicit and implicit simulations, the number of colloids $N_C$ is between 10 and 100. We assume that the ratio between the diameters of the colloids and that of the solute particles is $\sigma_C/\sigma_s=5$. The sizes of the colloid and the solute particles are  chosen to be of similar orders of magnitude for two reasons: (i) first, we believe that it is relevant to describe the formation of microemulsions in the intracellular media, which are typically made of very large macromolecules whose size is comparable to the other entities (organelles, enzymes, proteins...); (ii) second, we recently showed that propulsion could still be observed for larger ratios $\sigma_C/\sigma_s=10$, although with a larger computational cost~\cite{Decayeux2022}.

We simulate the trajectories of all particles (colloid and solute) using Brownian dynamics simulations~\cite{ErmakJCP75,Frenkel}.
The position $\rr_i$ of each of the $\Nc+N$ particles at time $t+\Delta t$ is determined from the positions at the previous step by the integrated overdamped Langevin equation: 
\begin{align}
    \vec r_i(t+\Delta t)  =& \rr_i(t)- D_{0,i} \sum_{i \ne j} \vec{\nabla} U(r_{ij}) \Delta t  \nonumber\\
    &+ \sqrt{2 D_{0,i} \Delta t }\vec {\eta}_i(t)
\end{align}
where $D_{0,i}$ is the diffusion coefficient of the particle $i$ at infinite dilution (or the `bare' diffusion coefficient), $\vec{\eta}_i$ is a random variable that follows a Gaussian distribution of mean equals to $0$ and of variance equals to $1$. By considering that this dynamics holds, we actually assume that both solute and colloids are sufficiently large to ignore inertial effects.

There are two types of solute particles in the system, named $A$ and $B$. $A$ particles interact with each other through the purely repulsive Weeks-Chandler-Andersen (WCA) potential~\cite{Weeks1971}:
\begin{equation}
\label{eq:def_WCA}
    U_{\text{WCA}}(r) =
    \begin{cases}
    4\varepsilon'\left[ \left( \frac{\sigma_s}{r}\right)^{12}-\left(\frac{\sigma_s}{r} \right)^{6}\right] + \varepsilon' & \text{if $r<2^{1/6}\sigma_s$}, \\
    0 & \text{otherwise},
    \end{cases}
\end{equation}
with $\varepsilon'=10k_{\rm B}T$, 
while $B$ particles interact through an attractive  Lennard-Jones (LJ) potential:
\begin{equation}
     U_{\text{LJ}}(r) =
    4\varepsilon\left[ \left( \frac{\sigma_s}{r}\right)^{12}-\left(\frac{\sigma_s}{r} \right)^{6}\right] .
\end{equation}
In order to improve computational efficiency, we neglect the long-range effects that may result from the LJ interactions, and impose a cutoff at a distance $2.5 \sigma_s$ for the $B$-$B$ interactions. The other pair interactions in this explicit model are as follows: $A$ and $B$ particles interact with each other through the WCA potential of Eq. \ref{eq:def_WCA}, colloids also interact with each other through this WCA potential with $\sigma_s$ replaced by $\sigma_C$, and colloids interact with solute particles of type $A$ or $B$ through the WCA potential with $\sigma_s$ replaced by ($\sigma_C$+$\sigma_s$)/2.  Hydrodynamic interactions between particles are not taken into account within this numerical scheme. Although such interactions were shown to affect the collective dynamics of self-propelled colloids { and may reduce clustering}~\cite{Matas-Navarro2014,Navarro2015,Zottl2014},  including them within our numerical study would be an important step forward, that we leave for future work.

The colloidal particles catalyze the reaction $A \to B$ in their vicinity, thus locally changing the nature of the interaction between solute particles. The reactions occur inside spherically symmetric reaction areas of radius $\rcut$, centered around each colloidal particle. Outside any reaction areas, the reaction  $B\to A$ takes place to ensure a non-equilibrium renewal of $A$ particles. We assume that both reactions occur very fast, in such a way that their characteristic time is smaller than all the other timescales of the simulation. Note that the $A\to B$ conversion rate is not enhanced when a solute particle belongs to several reaction areas simultaneously, compared to when it belongs to one reaction area only. For a system with one colloid, it has been shown that this mechanism leads to the self-propulsion of the colloid, characterized by an enhanced diffusion coefficient~\cite{Decayeux2021a, Decayeux2022}. 
The long-time diffusion coefficient of the colloid is computed from its mean-square displacement (MSD) as a function of time $\langle\Delta \rr^2_{\C}(t)\rangle =\langle(\rr_{\C}(t)-\rr_{\C}(0))^2\rangle$:
\begin{equation}D_{\rm eff}=\lim_{t \to \infty}\frac{\langle\Delta \rr^2_{\C}(t)\rangle}{4t},
\end{equation}
where the average is taken over multiple noise realisations. In every case, the MSD is computed at steady state. 

In this explicit model, the propulsion mechanism has been characterized in previous articles~\cite{Decayeux2021a,Decayeux2022}: for suitable parameters, the Lennard-Jones fluid demixes into a dense phase and a dilute phase, and liquid droplets form around the colloid. These mesoscale droplets stay confined in the reaction area close to the colloid surface (Fig. \ref{fig:snapshot}(a)), and tend to push the colloid. 
If the droplets orientation persists long enough, the colloid has an intermediate ballistic trajectory, and displays enhanced diffusion on a longer timescale  ~\cite{Decayeux2021a,Decayeux2022}. 
Here, we choose  parameters for which the self-propulsion is strong~\cite{Decayeux2021a,Decayeux2022}: the solute surface fraction is $\phi_s=0.079$, the intensity of the LJ attraction is $\varepsilon=3 \kB T$, which is sufficient to trigger phase separation if the local density of $B$ particles is large enough, and $\rcut=7.5 \sigma_s = 1.5 \sigma_C$. 
The surface fraction of colloids $\phi_{C}=\pi \sigma_{C}^2 \Nc / (4 \lbox^2)$ varies.

 \subsection{Implicit model}
\label{sec:model_ABP}

We now describe the implicit simulation scheme, on which we will map the explicit simulations. We rely on the ABP model,
where the dynamics of each colloid is described by the following dimensionless equations integrated with the Euler scheme:
\begin{align}
\label{ABP1}
   & \rr_i(t+\dt)  = \rr_i(t) +v_0 \nn_i(t) \dt \nonumber\\
   &- D_{t,C} \sum_{i \ne j} \vec{\nabla }V(r_{ij})\dt + \sqrt{2 D_{t,C} \dt}\vec{\xi}_i(t) ,\\
   \label{ABP2}
  &   \theta_i(t+\dt)  = \theta_i(t) + \sqrt{2D_{r,C} \dt} \xi_i'(t) .
\end{align}
 $\nn_i (t)=(\cos \theta_i(t), \sin \theta_i (t))$ is the orientation of the propulsion, whose persistence decays exponentially with a characteristic time $1/D_{r,C}$, in such a way that $\moy{\nn_i(0) \cdot \nn_j(t)}=\e^{-D_{r,C} t}\delta_{ij}$.  $D_{t,C}$  and $D_{r,C}$ are the translation and rotational diffusion coefficients, respectively. Both $\boldsymbol{\xi}_i$ and $\xi_i'$ are random variables which follow Gaussian distributions of zero mean  and of unit variance. We use reduced units to make the equations dimensionless: the distances are measured in units of $\sigma_\C$, the times in units of $\sigma_\C^2/D_{t,\C}$ (representing the time needed by a colloid to diffuse in a area of its diameter), and the energy  in units of $\kB T$. Finally, the potential $V(r)$ represents the interactions between the ABP particles.

 The implicit model [Eqs. \eqref{ABP1} and \eqref{ABP2}] therefore includes two main ingredients: the propulsion of individual colloids (characterized by the velocity $v_0$ and the rotational diffusion coefficient $D_{r,C}$) and the interactions between the colloids (characterized by the potential $V$). In what follows, we deduce these effective parameters from the simulations where the propulsion mechanism is described explicitly (Section \ref{sec:model_explicit}).

\section{Effective parameters for the implicit simulations}
\label{coarse-graining}

In this section, we calibrate the parameters of the implicit simulations. Under the assumptions of the ABP model, the active term $v_0 \nn_i(t)$ and the passive term $-D_{t,C} \sum_{i \ne j} \vec{\nabla }V(r_{ij})$  of the  equations of motions [Eqs. \eqref{ABP1} and \eqref{ABP2}] are independent. Therefore, we obtain each term in situations where the other one is absent. First, the parameters of the active term are derived from explicit simulations where colloid-colloid interactions are negligible. In other words, a single colloidal particle is placed in the simulation box, with $N$ solute particles and we run the explicit simulations described  in Section \ref{sec:model_explicit}.  Second, the effective potential describing the passive term is derived from explicit simulations without activity, i.e. when the colloid does not catalyze the $A\to B$ reaction, with only two colloidal particles in the simulation box so as to avoid $3$- or $N$-body contributions. 
    We use Monte Carlo simulations for this calculation.

\subsection{Effective activity parameters}
\label{sec:calibrating_activity}

 \begin{figure}
     \centering
       \includegraphics[width=0.7\columnwidth]{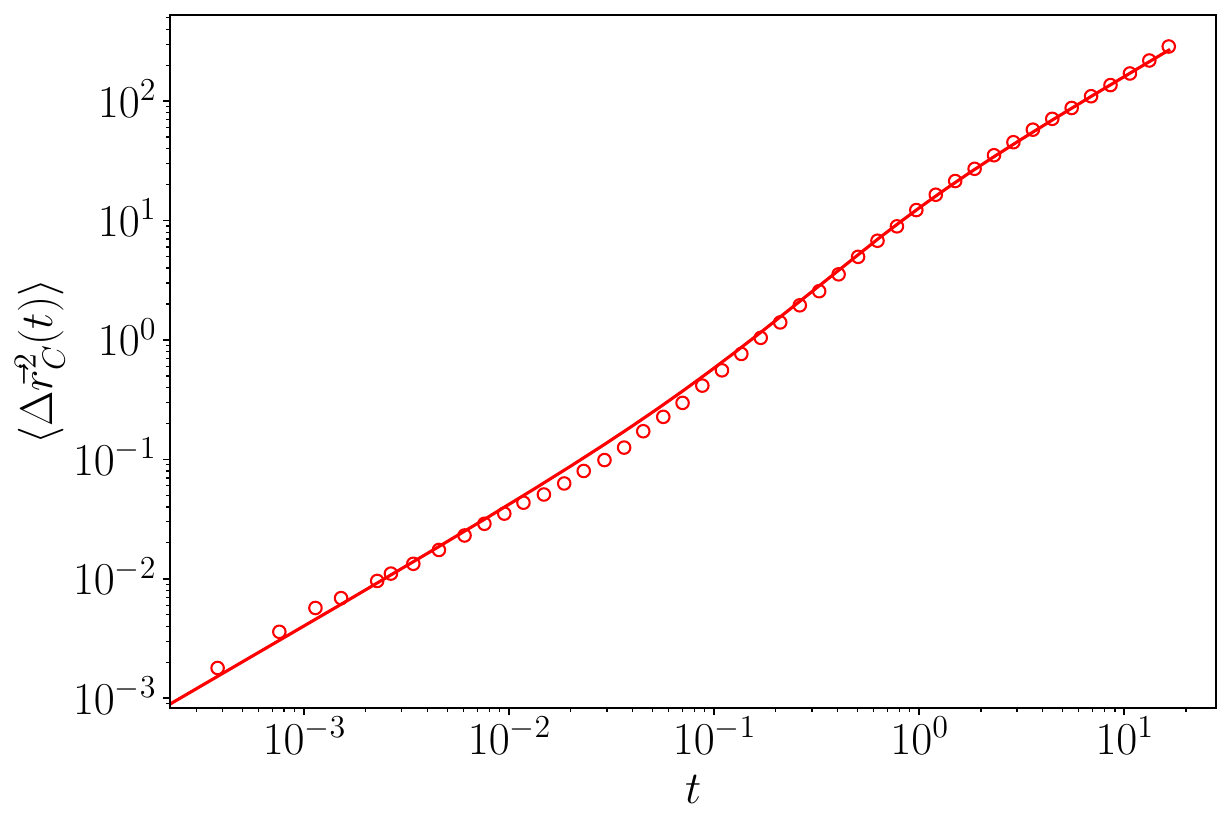}
    
\includegraphics[width=0.7\columnwidth]{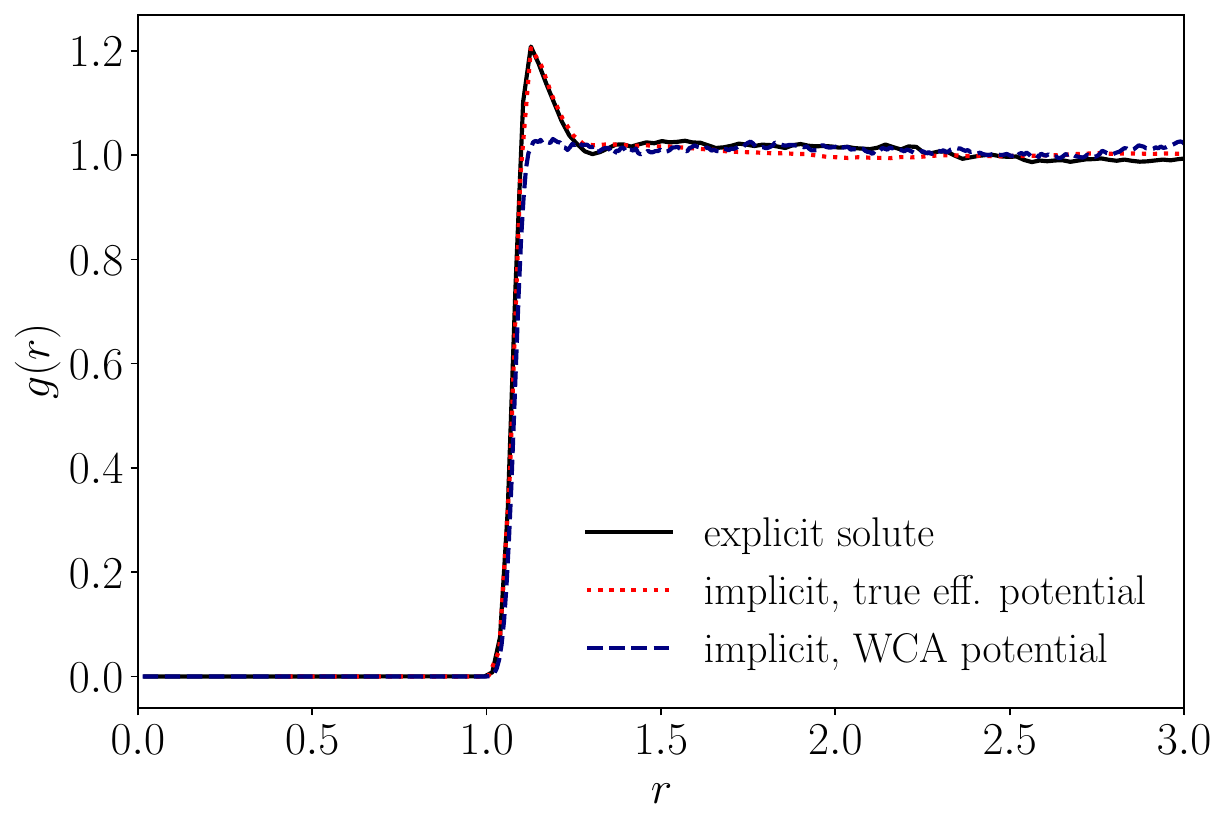}
     \caption{\textit{Top:} Symbols: Mean-square displacement of a single active colloid ($N_C=1$) as measured in the explicit simulations. Solid line: Mean-square displacement given by Eq. \eqref{eq_msd}, with the set of parameters $D_{t,C}$, $D_{r,C}$ and $v_0$ obtained from the explicit simulations. \textit{Bottom:} Radial distribution functions of passive colloids in the dilute limit ($\phi_C=0.01$) measured in explicit simulations and in the ABP simulations, for different colloid-colloid interaction potentials.}
     \label{fig:calibrating_ABP}

 \end{figure}

We first determine the parameters of the active term of the implicit simulations, which is  characterized by the magnitude of the propulsion velocity $v_0$ and the rotational diffusion coefficient of this velocity, $D_{r,C}$. The exact analytical expression of the MSD of a single ABP is~\cite{Howse2007}
\begin{equation}
\langle \Delta \rr_{\C}^2 (t) \rangle = 4  D_{t,\C} t + \frac{2 v_0 ^2}{D_{r,\C}} \left( t D_{r,\C} + \e^{-D_{r,\C} t} -1  \right).
\label{eq_msd}
\end{equation}

At short times,  this MSD becomes $\langle \Delta \rr_{\C}^2 (t) \rangle\sim 4  D_{t,\C} t$.  We thus deduce the translational diffusion coefficient $D_{t,\C}$ from a linear fit of the mean-square displacement at short time obtained by explicit simulations. 

We compute the rotational diffusion coefficient of the colloid $D_{r,C}$ from the explicit solute simulations. More precisely, as the propulsion of the colloid is related to the orientation of the solute cloud around it, we compute the  polarity vector $\vec{p}$, defined as $\vec{p}=\sum_{i\in\mathcal{P}} [\rr_i(t) - \rr_\text{C}(t)]$, 
where $\rr_i(t)$ is the position vector of solute $i$, and $\mathcal{P}$ is the circular area around the colloid of radius $(\sigma_C+3\sigma_s)/2$. We deduce the rotational diffusion coefficient from  the long-time limit of the autocorrelation function of the polarity vector:
$\moy{\pp (0) \cdot \pp(t) }\propto \e^{-D_{r,C}t}$~\cite{Decayeux2021a,Decayeux2022}. 
In the situation simulated with the explicit method (Section \ref{sec:model_explicit}), we find $D_{r,C}=3.24$ (in units of $\sigma_C^2/D_{t,C}$).

Finally, we extract $v_0$ from a fit of the MSD obtained by explicit simulations in the long-time limit, using the long-time limit of Eq. \ref{eq_msd}: $\moy{\Delta \rr_{\C}^2 (t) } \sim \left( 4 D_{t,\C} + \frac{2 v_0^2}{D_{r,\C}} \right) t $. We obtain $v_0=4.49$ (in units of $D_{t,C}/\sigma_C$). Fig.~\ref{fig:calibrating_ABP}(a) shows the MSD obtained from the explicit simulations, and the MSD given by Eq. \eqref{eq_msd} with the parameters we inferred.

\subsection{Effective interaction potentials}
Effective interactions between the colloidal particles differ from the direct WCA pair potential used in the explicit model, since the smaller solute particles create depletion interactions~\cite{Asakura1958}. 
The usual methods to obtain the interaction potential at finite concentration, such as HNC inversion~\cite{Bolhuis2002} or reverse Monte Carlo methods~\cite{Lyubartsev95}, are not directly applicable to non-equilibrium systems, since activity affects the structural and thermodynamical properties of the systems. Nevertheless, 
if one assumes independence between activity and effective interactions, the  effective potential $V(r)$ may be computed from  equilibrium simulations of the explicit solute model of a passive colloidal system.
This should be done at the same solute density as the active system, but without chemical reaction. 
Following a procedure first proposed by Wu and coworkers~\cite{Wu98}, we compute the mean force between the two colloidal particles. 
For a given distance between the colloidal particles, the forces are averaged over the configurations of the WCA solutes by using NVT Monte Carlo simulations (between 4$\cdot10^8$ and 6$\cdot10^8$ steps of length $1$). In these conditions, the two-body term of the $N$-body effective potential between the particles is derived. The colloidal particles are placed symmetrically along the diagonal of a cubic simulation box with periodic boundary conditions. The box size is sufficiently large to ensure that the solute distribution around both colloidal particles is not perturbed by the periodic images. 
We obtain a tabulated mean force that can be used as input effective force in standard Brownian dynamics or in implicit simulations. 

Fig. \ref{fig:calibrating_ABP}(b) displays the pair distribution functions between two colloidal particles obtained at equilibrium in a dilute system ($\phi_C=0.01$) from both explicit  and implicit simulations. 
We obtain a perfect agreement between explicit and implicit solute descriptions, as expected as we use the "true" effective interactions between colloids. The small peak at short distance is due to depletion effects from the solutes. As a matter of comparison, we also perform simulations with a simple repulsive WCA potential between colloids (with the size parameter $\sigma_C$ and an energy parameter $\varepsilon=10 k_{\rm B}T$). Of course, the simple WCA interaction potential  cannot account for the depletion effect, and no peak appears in the pair distribution function at this small surface fraction.

\section{Comparison of the models}
\label{section-structure}

\subsection{Structural properties: pair distribution functions}

\begin{figure}
     \centering
         \includegraphics[width=0.49\columnwidth]{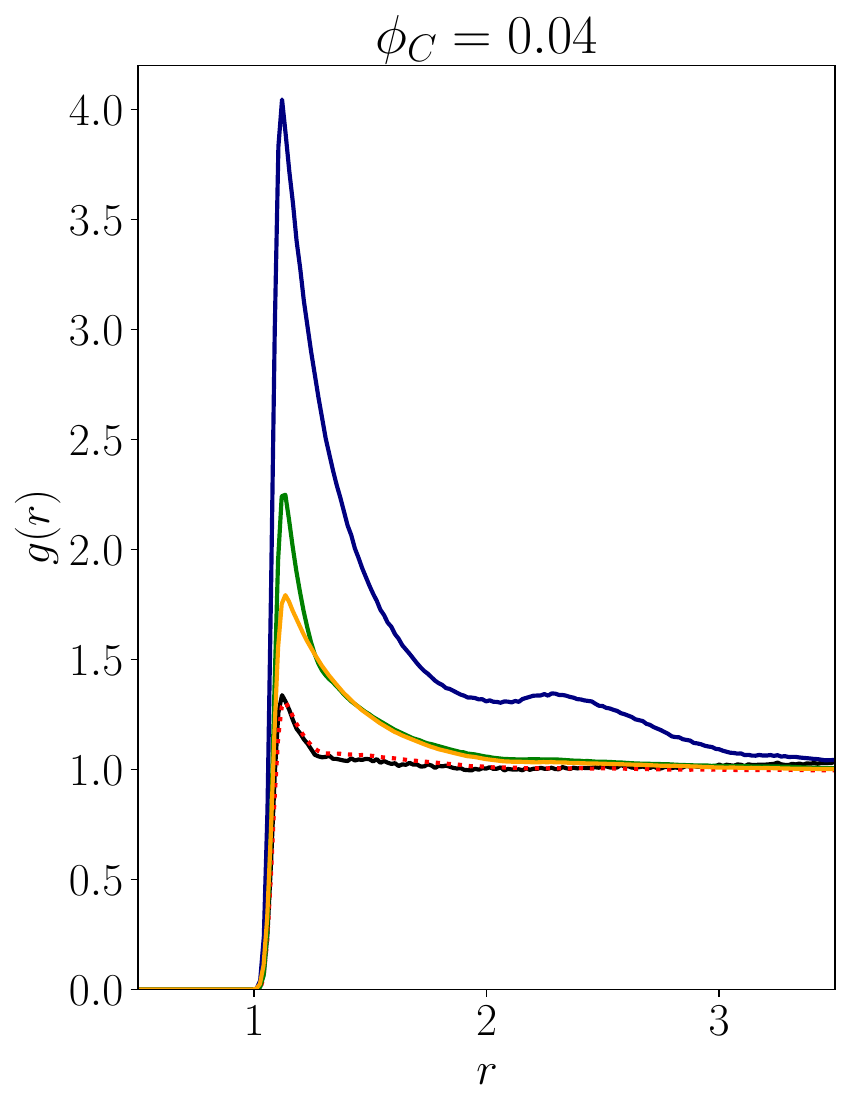}
     \includegraphics[width=0.49\columnwidth]{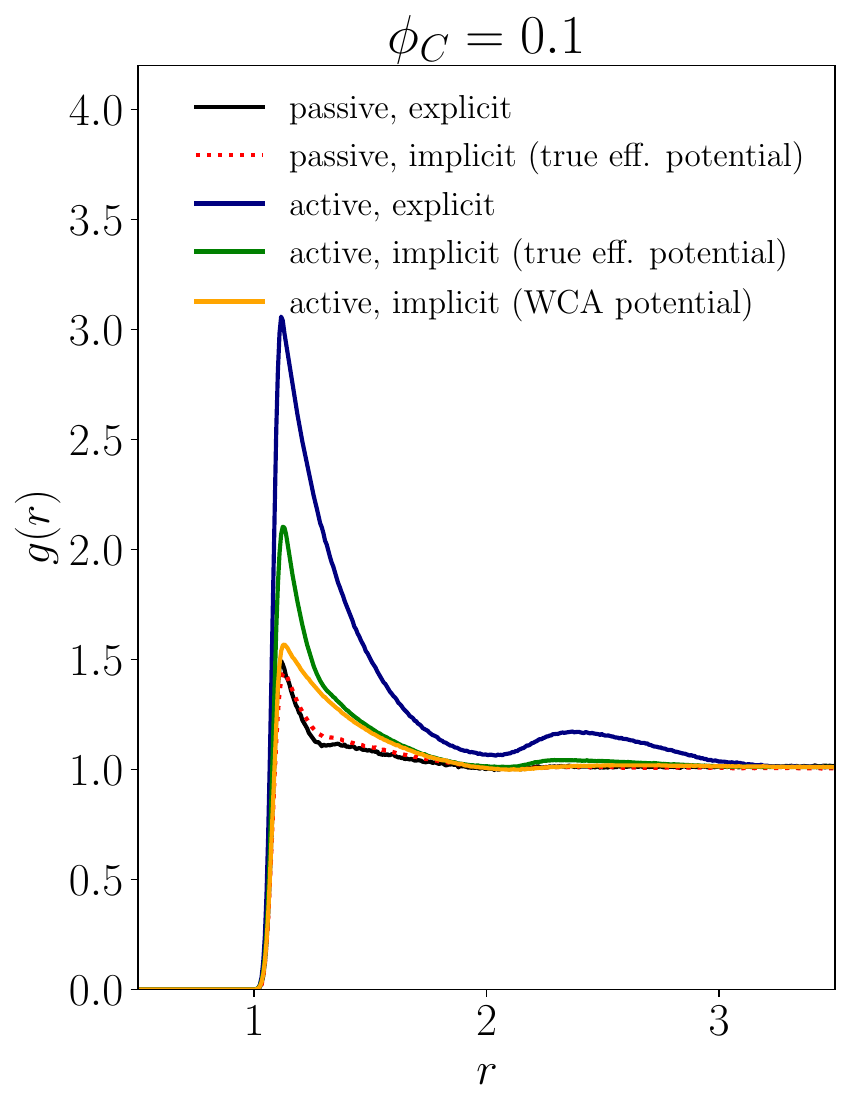}
     \caption{Radial distribution functions between colloidal particles for explicit solute and implicit solute models, at equilibrium or with the influence of a non-equilibrium reaction (active systems), for a surface fraction $\phi_C=0.04$ and $\phi_C=0.10$. Active implicit models are simulated through the Active Brownian Particle (ABP) algorithm, either using the exact infinite dilution effective potential, or using the same WCA potential as in explicit solute simulations. 
         }     
     \label{fig:gr_with_activity}
 \end{figure}

 We first check that the 
 effective force computed from Monte Carlo simulations accurately predicts the evolution of the structure with the colloidal density. 
We display  in Fig. \ref{fig:gr_with_activity} (black and red curves) the pair distribution functions obtained for two colloid surface fractions $\phi_C=0.04$ and $\phi_C=0.10$ from both explicit and implicit  simulations at equilibrium. 
There is  a remarkable agreement between both representations. It shows that, for such systems at equilibrium, there are no signature of n-body effects, {\em i.e.} the pair additive assumption for interactions between colloids performs perfectly well.

We then turn towards the analysis of non-equilibrium systems at steady state. 
The results are also displayed in Fig. \ref{fig:gr_with_activity}.
At these moderate colloid densities, the distribution functions obtained from explicit simulations exhibit two broad peaks. Those obtained from implicit simulations only present a single visible peak. 
Such a peak at short distance is the signature of an effective attraction between colloidal particles. The broadness of the first peak and the presence of a second one in the results obtained by explicit simulations are the sign of a strong structuring of the suspension at steady state. In all cases, the first peak is much higher than that induced by depletion at equilibrium. However, the impact of the non-equilibrium ingredient -- the activity of ABP for the implicit solute model, and the chemical reaction in the explicit solute model -- is significantly greater for the explicit solute simulation.
As a matter of comparison, we perform implicit simulations  with a simple WCA interaction potential between colloids instead of the exact infinite dilution potential.  We see a change of the radial distribution function at short range, but the long-range part of the structure is not significantly affected by the modification of the interaction potential (see the green and orange lines in Fig. \ref{fig:gr_with_activity}). The influence of the interaction potential used in ABP simulations is thus considerably less important than the difference between implicit and explicit descriptions of the solutes. 

This structuring of the suspension was expected as activity is known to induce the formation of clusters, and even to cause phase transitions for ABP models~\cite{Fily2012,Redner2013}.  However, we are here in a range of densities and propulsion parameters for which the association of active colloids stays moderate. We made a cluster analysis from the trajectories, for both explicit and implicit simulations. We assume that a cluster is an assembly of two or more particles that are at distances smaller than $d_{\rm cutoff}=1.5 \sigma_C$. The size of clusters is computed from positions at steady-state sampled at a frequency large enough to ensure that two configurations of the same system are independent. In the systems investigated here, colloidal particles mainly tend to associate into pairs, sometimes into triplets, and very rarely into larger clusters. 

The probability for a colloidal particle to be in a pair or in a triplet is computed for both implicit and explicit models and shown in Tab. \ref{table-cluster}. We show again that the effective model perfectly predicts structural properties for equilibrium (passive) systems, as the agreement with the explicit solute model for both pair and triplet probabilities is excellent. 
However, we confirm that the effective model strongly underestimates the association of active colloidal particles into pairs or triplets. In our systems, the probability to have pairs in the explicit solute model is roughly twice larger than that of the implicit model, for all packing fractions below $\phi_C=0.05$. The probability of triplets is much larger in the explicit solute case, from roughly $50 \%$ higher for $\phi_C=0.10$ to more than five times higher at $\phi_C=0.01$.

For the highest colloidal density, $\phi_C=0.10$, the difference in the pair probabilities for implicit and explicit models is small. The low sensitivity of the structure to the modeling of attractive interactions is a common behavior at high densities. It can be understood in the limit where the colloid density tends towards that of a colloidal liquid where particles are closely packed. In the latter case, the effective attractions are no longer influencing the pair distribution functions~\cite{Hansen1986}, since the particles cannot get any closer from each other. Nevertheless, the triplet probabilities remain quantitatively different, and show that even at this density, the explicit description of solute-induced forces leads to more clustering.

\begin{table}
\begin{center}
\begin{tabular}{|c | c |c |c |c |}
\hline
surface  &  state & solute   & pair  & triplet  \\
fraction $\phi_C$ &   &  model  & probability &  probability  \\ \hline
{$0.01$}& passive & impl. &  0.028 & 0.000\\
 & & expl. &  0.027 & 0.000 \\
 & active & impl. &  0.042 & 0.002 \\
 &  & expl. & 0.098 & 0.011\\ \hline \hline
$0.03$& passive & impl. &  0.083 & 0.008 \\
 & & expl. &  0.081 & 0.008 \\
 & active & impl. &  0.117 & 0.016 \\
 &  & expl. & 0.186 & 0.056\\ \hline \hline
$0.04$& passive & impl. &  0.105 & 0.012\\
 & & expl. &  0.102 & 0.012 \\
 & active & impl. &  0.128 & 0.023 \\
 &  & expl. & 0.196 & 0.060 \\ \hline \hline
 {$0.10$}& passive & impl. &  0.199 & 0.060\\
 & & expl. &  0.195 & 0.061 \\
 & active & impl. &  0.218 & 0.088 \\
 &  & expl. & 0.223 & 0.114 \\ \hline \hline
\end{tabular}
\end{center}
 \caption{Steady-state probabilities for two or three colloidal particles to be associated into pairs or into triplets, for equilibrium (passive) systems and active ones; for implicit (or effective) solute models and for explicit solute models.}
\label{table-cluster}
\end{table}

\subsection{Dynamical properties: effective diffusion coefficients}
\label{section-dynamics}

  \begin{figure}
     \centering
     \includegraphics[width=1\columnwidth]{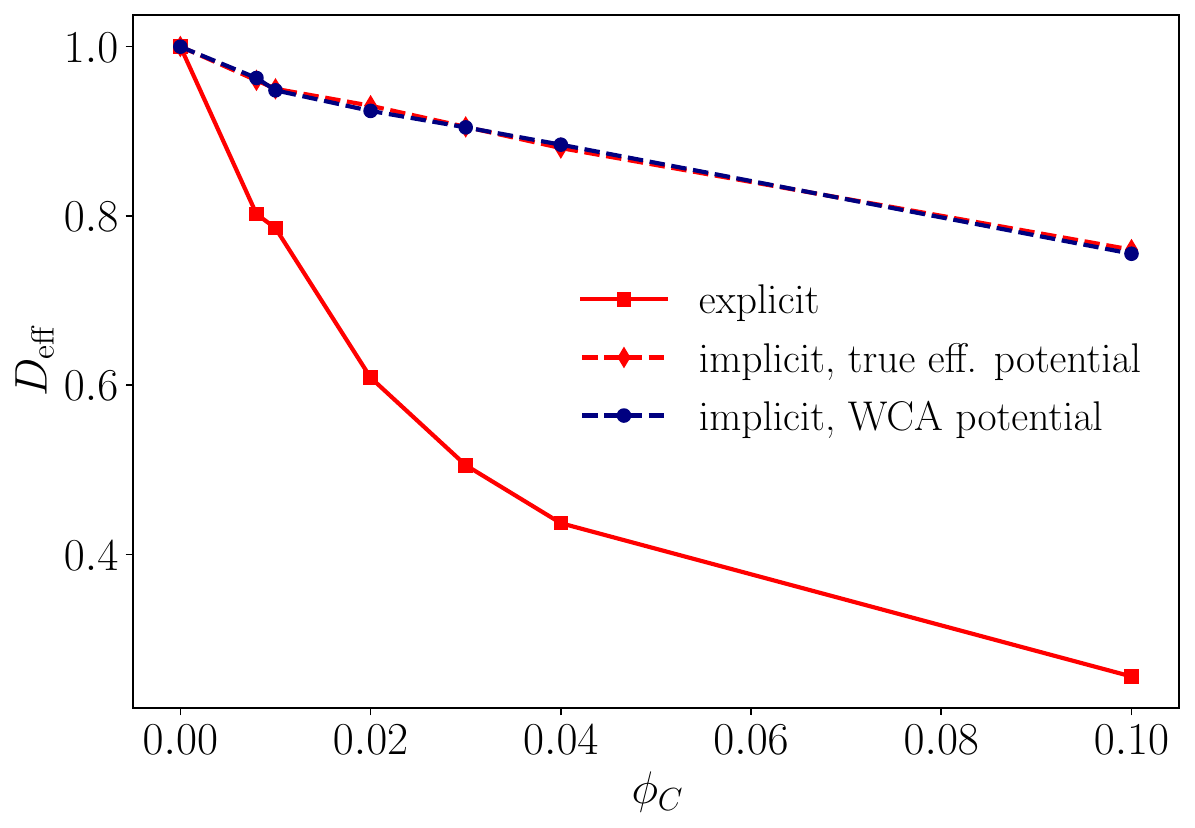}
     \caption{Effective diffusion coefficient of the colloids as a function of their surface fraction $\phi_C$, computed both in the explicit model and with the implicit ABP simulations with different interaction potentials. }
     \label{fig:effective_diffusion_coefficients}
 \end{figure}

 The dynamical properties are analyzed through the computation of the mean-square displacement of the colloids (MSD) as a function of time. The average is done over multiple noise realisations and over all the colloidal particles. 
A series of simulations for varying colloid densities shows that the long-time diffusion coefficient progressively decreases with the colloid surface fraction (Fig~\ref{fig:effective_diffusion_coefficients}).
This behavior is seen with both implicit and explicit  models. 
In other words, self-propulsion is less effective in enhancing the diffusion coefficient when the colloidal density increases. 
Nevertheless, the dynamical quantities are significantly less affected by the colloidal density in the implicit  model than in the explicit  model. 
It also implies that enhanced diffusion is more important for the ABP model than for the explicit solute model at all densities.  
For instance, the decrease of the diffusion coefficient for $\phi_C = 0.10$ is of $75 \%$ for the explicit solute model and of at most $25 \%$ for the implicit solute model. 
Also, it should be noted that the effective diffusion coefficient of ABP particles is almost not affected by the choice of the effective interaction potential. Indeed, replacing the infinite dilution effective potential by a mere WCA potential does not affect significantly the diffusion coefficient, as the difference is of the same order of magnitude as the uncertainty of the calculation, which is smaller than $1 \%$.

\section{Conclusive discussion}

Both structural and dynamical properties of the implicit and explicit models show qualitative differences. In the implicit ABP model, the influence of activity on the colloidal structure is well known. The colloid-colloid position correlations (that can be read in the radial distribution function)  generally increase with activity, because active particles tend to propel into neighbouring particles. This illustrates a key coupling between dynamics and structure in implicit solute models of active particles: the faster the particle goes, the more likely it clusters with other particles.  
This leads to an apparent contradiction with our explicit solute model: at a given finite density, the explicit model predicts that particles are slower than in ABP, which is a dynamical signature of a reduced activity. 
Conversely, these explicit simulations
display higher structural correlations, as it can be seen in the radial distribution functions, which is a structural signature of a higher activity.

 The aforementioned contradiction prevents the use of density-dependent effective parameters to accurately map the properties of the explicit model into an implicit one. 
If one keeps the infinite dilution effective potential and infers density-dependent activity parameters, then 
activity must be lowered to fit the transport coefficients. In that case, since lowering activity decreases clustering, the quality of structure prediction would decrease. Therefore, a density-dependent activity would not be a solution. If one keeps the effective activity parameters constant and infers density dependent effective potentials, no more success is expected, since we have shown that the impact of the effective potential on transport is very small.  
Indeed, when we have tried to compare the model with a true effective potential to a mere WCA potential with no effective attraction at all, the impact on transport was not significant. Therefore, density-dependent effective parameters would not be sufficient to map explicit models onto implicit ones.

This important divergence of explicit and implicit solute simulations comes from the complex coupling between solutes in the explicit model. 
{ When the colloid density increases, 
the number of droplets per colloidal  particle decreases, as shown in the Appendix section. 
Actually, as it can be seen on typical snapshots from the explicit simulations (see Appendix), droplets are constrained outside of the clusters of colloids. This is corroborated by the fact that clusters of colloids are more stable in the explicit model than in the implicit one, as droplets keep pushing colloids towards the center of the cluster. This non-trivial coupling between the structure of the solute cloud and the configuration of the colloidal particles partly explains the failure of the implicit solute model.}

Our study shows that, when solute-solute interactions become a core ingredient of active phenomena, classical models of active Brownian particles  have to be  revisited. 
The use of coarse-graining strategies based on infinite dilution effective parameters, that were designed for passive colloids and successfully tested~\cite{Wu98}, does not work for our model of self-propelled particles. This observation may apply to other systems close to phase transitions, as most biological systems, for which very strong many-body effects limit the transferability of effective parameters to finite concentrations. 
Despite this apparent failure of active Brownian particle models, the comparison with explicit solute models unravels important aspects. Our explicit model is much more sensitive to concentration effects than the usual models of active matter. It means that the properties of the colloidal particles may be highly dependent on their environment, which opens many fundamental questions concerning active processes in biology. Cellular media are highly heterogeneous, and some biological macromolecules or assemblies may be active in subparts of the cells, and passive in other parts.

In summary, this work sheds light on the complexity of active systems driven by interactions within the surrounding bath. It naturally opens the way to the exploration of larger systems and stronger activities, where enhanced diffusion itself leads to phase separation. Then, the interplay between phase separations occurring at both solute and colloid scales may generate a vast variety of original dynamical and structural behaviors.
The importance of many-body effects suggests that, for active colloids, one may look at more sophisticated coarse-graining methodologies that have been set-up recently for highly polarizable passive systems. For passive colloids, recent works open the way to the use of machine learning to derive many-body potentials~\cite{Campos21}, which may be adapted to active colloids in future works.

\appendix

\section{Size distribution of droplets in explicit solute simulations}
\label{section-droplet}

\begin{figure}
    \centering
    \includegraphics[width=\columnwidth]{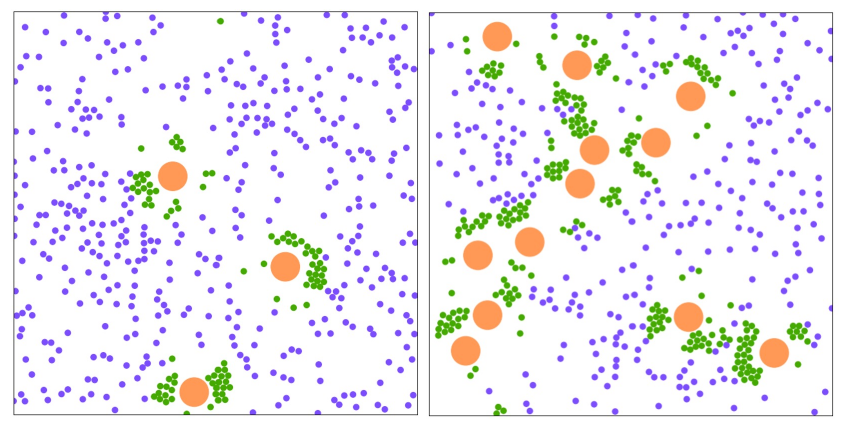}
    \caption{Typical snapshots of the system obtained from explicit solute simulations at stationary state. The colors are the same as in Fig. \ref{fig:snapshot}. The surface fraction of colloids is $\phi_C=0.012$ on the left, $\phi_C=0.048$ on the right.}
    \label{fig:snapshot2}
\end{figure}

In the explicit solute model, the Lennard-Jones interactions drive the formation of droplets of  solute particles close to colloidal particles, where the reaction converts $A$ particles (WCA interaction potential) into $B$ particles (LJ interaction potential). The growth of these droplets is limited by the conversion of $B$ particles into $A$ particles far from the colloids. We display in Fig. \ref{fig:snapshot2} typical snapshots obtained from simulations at the stationary state. We observe a certain size polydispersity of the droplets, which are located outside clusters of colloids when they exist. We have analyzed the size distribution and the number of droplets for several surface fractions of colloids. The size of the simulation box is in any case $\lbox=70\sigma_s$, and the surface fraction of colloids is varied between $\phi_C=0.012$ and $\phi_C=0.096$ by changing the number of colloids in the simulation box. The analysis of the structure of the system is performed at stationary state, and the results are averaged over $75$ independent trajectories.

\begin{figure}
    \centering
    \includegraphics[width=\columnwidth]{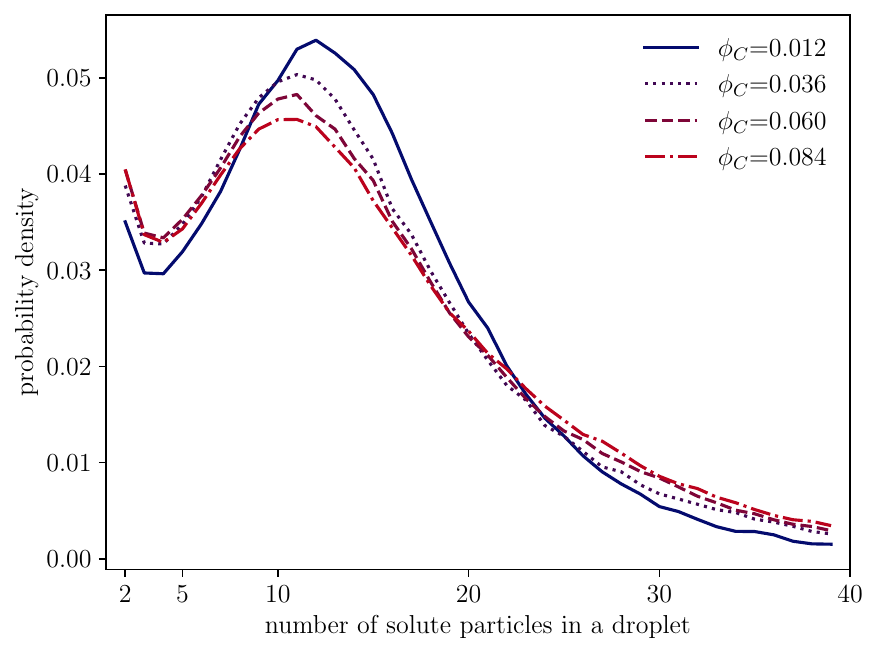}
    \caption{Probability density of having a given number of $B$ solute particles (Lennard-Jones particles) in a droplet for several surface fractions of colloids $\phi_C$.}
    \label{fig:droplet_size}
\end{figure}

\begin{figure}[b]
    \centering
    \includegraphics[width=0.49\columnwidth]{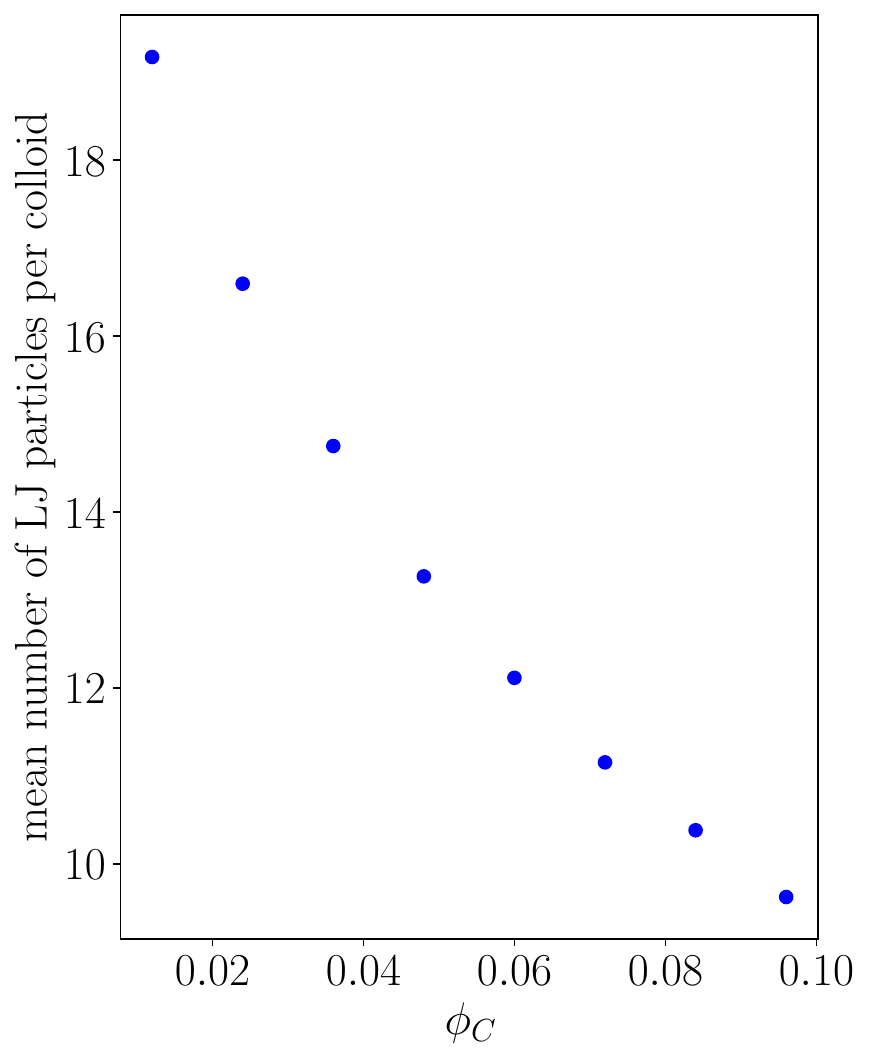}
    \includegraphics[width=0.49\columnwidth]{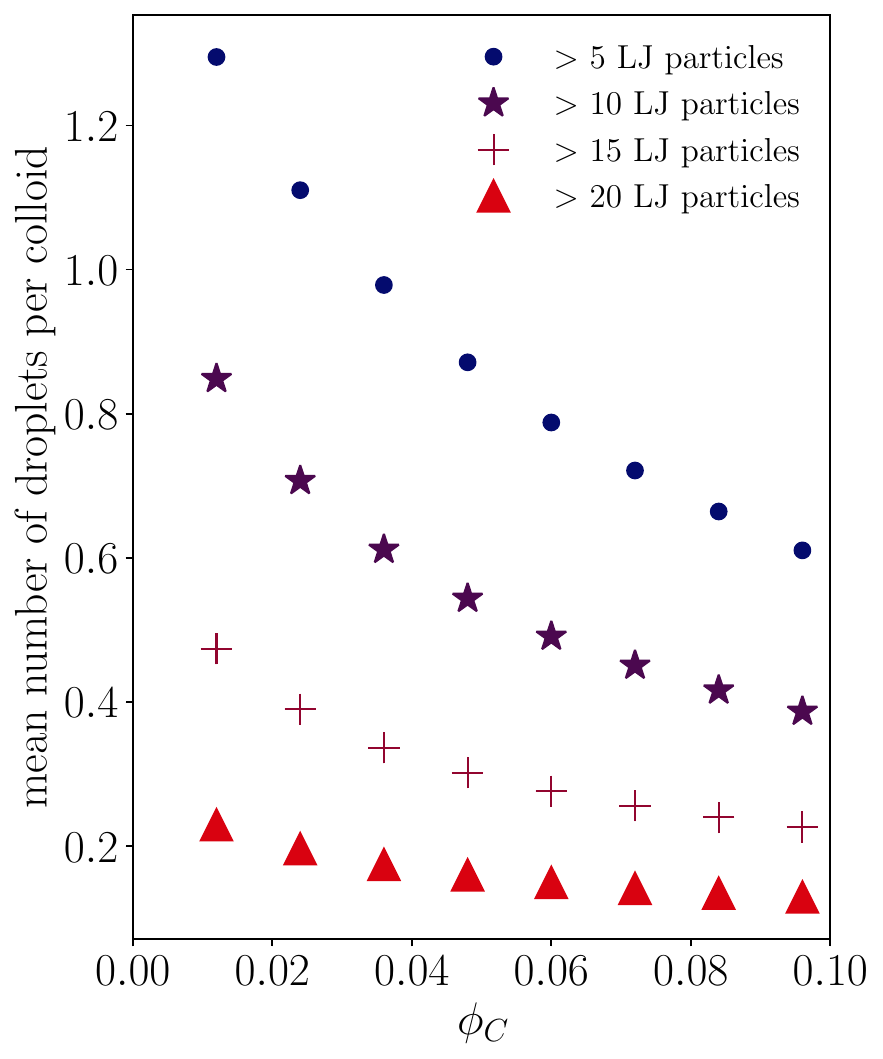}
    \caption{(left) Mean number of $B$ solute particles (Lennard-Jones particles) per colloid as a function of the colloid density $\phi_C$. (right) Mean number of droplets per colloid as a function of the colloid density $\phi_C$. Different plots correspond to different definitions of a droplet, that is assumed to contain a number of solute particles larger than a threshold. The threshold varies between $5$ and $20$ solute particles in a droplet.   }
    \label{fig:droplet_count}
\end{figure}

To characterize the number and the size of droplets of solute particles, we use a Voronoi cell analysis \cite{Lazar2022}.  First, the size distribution of the Voronoi cells was computed for solute particles.  The formation of droplets of $B$ particles leads to a bimodal distribution of the size of the Voronoi cells for solutes, that allows us to define a threshold below which a particle can be tagged as being part of a droplet. The threshold value is chosen equal to $2\sigma_s$.
The density probability of the number of particles per droplet is displayed in Fig. \ref{fig:droplet_size} for several surface fractions of colloids. This probability density presents a maximum that is shifted to smaller droplet sizes when the density of colloid increases. We compute the mean size of droplets from these probability densities, assuming that a droplet consists at least in $5$ solute particles and neglecting the smaller clusters. The mean size is found almost independent of the colloid density, and varies between $12.0$ and $12.5$ solute particles per droplet.

 We have also computed the average number of $B$ particles per colloid. The results are displayed in Fig. \ref{fig:droplet_count} (left).  This quantity is found to decrease almost by a factor two when the colloid density increases from $\phi_C=0.012$ to $\phi_C=0.096$. It is larger than the mean size of droplets at low densities but smaller than it at large densities. Also, the mean number of droplets per colloids is found to decrease with the colloid density whatever the threshold chosen to define a droplet. Results are displayed in Fig. \ref{fig:droplet_count} (right) for several thresholds used to define a droplet. It appears that the number of droplets per colloids is much smaller than $1$ when the colloid density increases, especially if we assume droplets contain at least $10$ particles. This proves that colloids share droplets when the colloid density increases.   
There is therefore a clear coupling between the number of colloids and the structure of the solute cloud around each colloid.

\end{document}